\documentclass[a4paper,UKenglish,cleveref, autoref, thm-restate]{lipics-v2021}
\nolinenumbers

\usepackage{listings}
\usepackage{tcolorbox}
\usepackage{xcolor}
\usepackage{tikz}
\usepackage{tabularx}
\usepackage{array}
\usepackage{booktabs}
\usepackage{threeparttable}
\usepackage{url}
\usetikzlibrary{positioning,arrows.meta,shapes.multipart}

\usetikzlibrary{arrows.meta,positioning}
\definecolor{delim}{RGB}{20,105,176}
\definecolor{numb}{RGB}{106, 109, 32}
\definecolor{string}{rgb}{0.64,0.08,0.08}
\usepackage{booktabs}
\usepackage{array}
\usepackage{ragged2e}
\newcolumntype{P}[1]{>{\RaggedRight\arraybackslash}p{#1}}

\lstdefinelanguage{json}{
    showspaces=false,
    showtabs=false,
    breaklines=true,
    postbreak=\raisebox{0ex}[0ex][0ex]{\ensuremath{\color{gray}\hookrightarrow\space}},
    breakatwhitespace=true,
    basicstyle=\ttfamily\small,
    upquote=true,
    morestring=[b]",
    stringstyle=\color{string},
    literate=
     *{0}{{{\color{numb}0}}}{1}
      {1}{{{\color{numb}1}}}{1}
      {2}{{{\color{numb}2}}}{1}
      {3}{{{\color{numb}3}}}{1}
      {4}{{{\color{numb}4}}}{1}
      {5}{{{\color{numb}5}}}{1}
      {6}{{{\color{numb}6}}}{1}
      {7}{{{\color{numb}7}}}{1}
      {8}{{{\color{numb}8}}}{1}
      {9}{{{\color{numb}9}}}{1}
      {\{}{{{\color{delim}{\{}}}}{1}
      {\}}{{{\color{delim}{\}}}}}{1}
      {[}{{{\color{delim}{[}}}}{1}
      {]}{{{\color{delim}{]}}}}{1},
}

\newif\iflongversion

\NewDocumentEnvironment{longversion}{ +b }{\iflongversion#1\fi}{}

\title{Prompts Blend Requirements and Solutions: From~Intent~to~Implementation}

\titlerunning{The Prompt Triangle}

\author{Shalini Chakraborty}{University of Bayreuth, Germany}{shalini.chakraborty@uni-bayreuth.de}{0000-0002-9466-3766}{}

\author{Jan-Philipp Steghöfer}{XITASO GmbH IT and Software Solutions, Augsburg, Germany}{jan-philipp.steghoefer@xitaso.com}{0000-0003-1694-0972}{}

\authorrunning{Chakraborty and Steghöfer}
\Copyright{Shalini Chakraborty and Jan-Philipp Steghöfer}


\begin{document}

\maketitle


\begin{abstract}
AI coding assistants are fundamentally reshaping software development by shifting developers' effort from writing code toward specifying intent through natural language prompts. In emerging chat-based development practices such as \emph{vibe coding}, prompts mediate the transformation of human intent into executable software. While Requirements Engineering (RE) emphasizes capturing, validating, and evolving requirements, current prompting practices remain informal and ad hoc.
In this vision paper, we argue that prompts represent lightweight, evolving requirements artifacts that combine expressions of user needs with varying degrees of solution guidance. We use an existing conceptual model that decomposes prompts into three interrelated dimensions: \emph{Functionality and Quality} (capturing intended system requirements), \emph{General Solutions} (capturing architectural strategies and technology choices), and \emph{Specific Solutions} (capturing implementation-level constraints and directives). 
Building on this conceptualization, we formulate four research hypotheses concerning (i) the evolution of prompts over time, (ii) the influence of user characteristics on prompt evolution, (iii) the relationship between prompt content and requirements validation and verification activities, and (iv) the impact of prompt characteristics on requirements and resulting software quality. We envision an empirical research agenda combining real-world AI-assisted development data, corpus analysis, and controlled experimentation to investigate these hypotheses and derive evidence-based practices for requirements-aware prompt engineering. By reframing prompts through the lens of RE, we position prompting not merely as an interaction mechanism with AI systems, but as a central software engineering concern requiring systematic study.

\end{abstract}

\section{Introduction}
AI-assisted chat-based coding tools fundamentally reshape software development by enabling developers to express functionality, constraints, and behavior in natural language rather than code. In emerging paradigms like vibe coding~\cite{ray2025review} and agentic coding~\cite{wang2025ai,robeyns2025self}, natural language prompts become the primary artifact bridging human intent and executable software. Unlike traditional workflows that maintain clear divisions between requirements, design, and implementation, chat-based development collapses these boundaries: a single prompt encapsulates requirements, architectural strategy, and implementation constraints simultaneously.

Yet prompts remain informal, ephemeral, and ad hoc, lacking the rigor traditionally associated with Requirements Engineering (RE). We argue that prompts should be understood as lightweight, evolving requirement artifacts that blend requirements with solution guidance, and that systematic prompt engineering represents a new frontier for RE in the age of generative AI.

This paper builds on the existing prompt triangle~\cite{shalini2026exploring} as an emerging result, a conceptual model decomposing prompts into three interrelated dimensions: (1) Functionality and Quality (the requirement), (2) General Solutions (architectural strategy and technology choices), and (3) Specific Solutions (implementation-level constraints). 
We validate this model using the DevGPT dataset~\cite{xiao2024devgpt}, position it relative to existing prompt patterns and RE validation techniques, and formulate four testable hypotheses about prompt evolution, user characteristics, requirements verification in iterative prompting, and the relationship between prompt refinement and code quality. Our vision is to establish empirically-informed best practices for requirements-aware prompt engineering, rejuvenating RE for AI-assisted development.

\section{Chat-based Coding Assistants}

AI coding assistants operate in two distinct modes that shape developer interaction fundamentally differently. \emph{Code completion interfaces} work within existing code structures, assisting with boilerplate, method bodies, and idiomatic patterns where developers provide the architectural skeleton. \emph{Chat-based interfaces} enable developers to describe functionality, constraints, and solution strategies in natural language, potentially revamping entire codebases from a single prompt.

This distinction is methodologically critical~\cite{barke2023grounded,vaithilingam2022expectation,sergeyuk2025using}: code completion preserves traditional incremental programming, while chat-based prompting externalizes problem-solving into conversational exchanges where a single prompt conflates requirements specification, design rationale, validation-verification criteria, and implementation intent. Conflating the two modes risks masking key differences in cognitive effort, developer mental models, and the role of requirements~\cite{wei2022chain}. This paper focuses on chat-based interfaces and their implications for requirements engineering.

\section{Prompts and Requirements}
Prompt engineering refers to the practice of carefully crafting and refining prompts to guide large language models (LLMs) toward producing desired outputs~\cite{reynolds2021prompt,chen2023unleashing}. In the context of software development, prompt engineering is increasingly seen as a crucial skill for effectively leveraging chat-based coding assistants. However, while prompts may contain elements of system requirements or design intent, they are often ad hoc and lack the structure of formal requirements~\cite{barke2023grounded}. Many developers struggle to provide the right level of detail and to translate requirements into prompts consistently~\cite{VillamizarFKVM25}. This raises an important question for RE: \textbf{How can the rigor of RE be infused into prompting practices?} Viewing prompts as lightweight but evolving requirement artifacts offers a way to bridge the gap between informal developer intentions and structured requirements specification~\cite{vogelsang2024prompting}. Such an integration will enhance the reliability of AI-generated solutions and redefine RE practices in the era of AI-assisted chat-based development.


\subsection{A Conceptual Model of Prompts}
\label{sec:prompts-requirements:conceptual-model}

Recent work has begun to examine the role of prompts in software development and Requirements Engineering (RE)~\cite{ronanki2025prompt,huang2025prompt}. However, understanding prompts as artifacts within AI-assisted development remains an open challenge. This is particularly relevant in emerging practices such as \emph{vibe coding}, where software is iteratively developed through conversational interactions with AI~\cite{meske2025vibecodingreconfigurationintent}, and \emph{agentic coding}, where autonomous AI agents reason about and execute development tasks, guided by the developer and a large language model (LLM). In such settings, prompts become the primary medium through which developer intent is communicated and translated into executable systems. Consequently, prompts increasingly occupy a role that extends beyond simple instructions and intersects with core concerns of RE, such as requirements elicitation, specification, and evolution.

\begin{figure}[tb]
    \centering
    \includegraphics[width=.8\columnwidth]{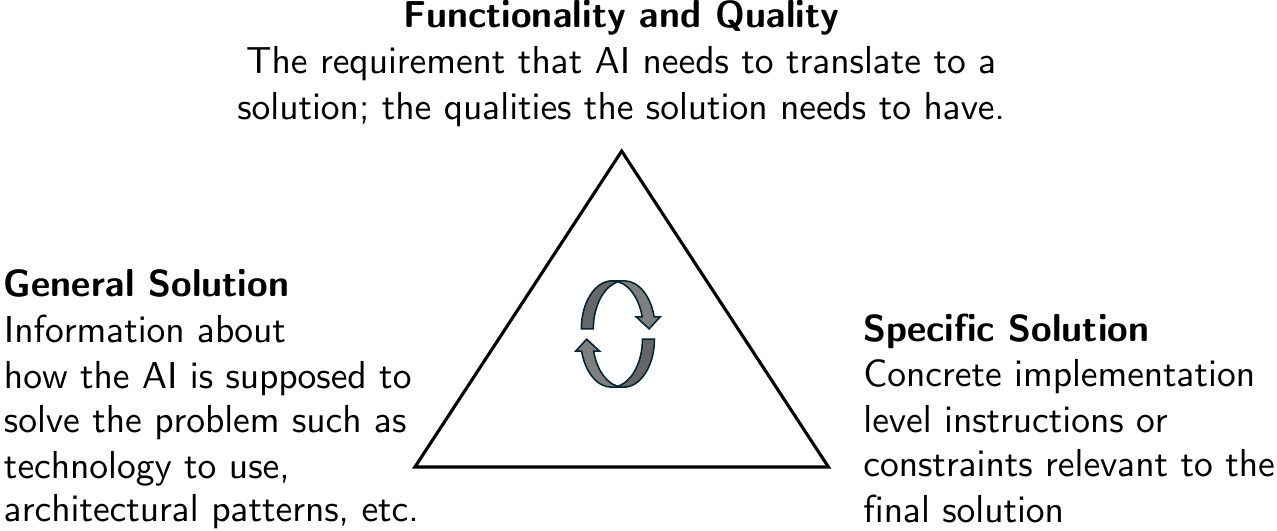}
    \caption{Prompt Triangle: prompts consist of three interrelated dimensions: Functionality and Quality, General Solutions, and Specific Solutions. The circular relationship indicates iterative refinement across development interactions.}
    \label{fig:promptTriangle}
\end{figure}

To better understand this role, we draw on an existing conceptual model that decomposes prompts into three interrelated components. The \emph{Prompt Triangle} (Figure~\ref{fig:promptTriangle}) provides a practical and conceptual lens for structuring prompts and assessing their completeness from an RE perspective. We posit that most prompts used in chat-based coding contain these three dimensions implicitly.

The \emph{Prompt Triangle} consists of three interrelated dimensions. \emph{Functionality and Quality} captures intended system behavior and desired quality characteristics, i.e., what the system should do and how well it should perform, encompassing both functional requirements and quality requirements such as performance, security, accessibility, reliability, and usability. \emph{General Solutions} capture higher-level guidance on how functionality should be realized, including architectural decisions, preferred technologies, design paradigms, and development strategies. \emph{Specific Solutions} represent detailed implementation instructions and constraints that further narrow the solution space and guide concrete realization.
Embedding quality attributes directly within the functionality description ensures that prompts capture not only \emph{what} a system should achieve, but also \emph{how well} it should achieve it. Whenever possible, these requirements should be expressed in a measurable way, while guidance on realization mechanisms remains within the General and Specific Solutions dimensions. Co-locating quality requirements with functionality stabilizes developer intent and creates a clearer foundation for validation and verification activities.

Unlike traditional requirements artifacts, which often separate requirements from implementation concerns, prompts frequently blend these dimensions within a single interaction. During iterative development, prompts evolve by expressing intent, refining solution strategies, and constraining implementation details. The circular relationship within the Prompt Triangle represents this continuous refinement process, where movement across components occurs as developers and AI systems progressively converge toward a desired solution.

As an \textbf{illustrative example} of the Prompt Triangle, consider the following prompt:

{\fontfamily{cmss}\selectfont
Build a React web app that shows the current weather for a given city, and make it fast, reliable, safe, and accessible. Use server-side rendering; fetch data from OpenWeather. Style the page cleanly and show city, condition, humidity, wind, and Celsius temperature with an icon. It should load quickly on a typical phone, show a loading state, retry if the network fails, and display clear errors. Keep the API key on the server, avoid logging personal details, and ensure people can use the app with a keyboard and screen readers with readable colors.
}

Although the prompt appears as a single instruction, it combines multiple concerns that can be mapped to the dimensions of the Prompt Triangle. \emph{Functionality and Quality} includes the requirement to build a weather application, together with quality characteristics such as performance, reliability, security, and accessibility (e.g., fast loading, robust error handling, privacy-preserving logging, and support for keyboard and screen-reader interactions). \emph{General Solutions} guide broader implementation decisions, including the use of React, server-side rendering, and integration with the OpenWeather API. Finally, \emph{Specific Solutions} further constrain implementation through concrete design choices such as displaying weather attributes (city, humidity, wind, temperature, icons), applying a clean visual style, and ensuring that API keys remain securely stored on the server.

We believe that this triangular model provides a lens through which we can examine prompts not just as one-off commands, but as \textbf{evolving artifacts of requirements expression, negotiation, and realization} which capture the main intentions of the developer~\cite{kruger2023intentions}. As prompts are iteratively refined, attention may oscillate between specifying what is needed (Functionality and Quality), how it should be approached (General Solutions), and how exactly it should be executed (Specific Solutions). The developer thus incrementally reshapes the solution space. In this sense, even prompts focused on implementation remain grounded in a shifting understanding of \emph{Functionality and Quality}.
%
%


\textbf{Model Validation}
Before devising an empirical research agenda around this model, we conducted an initial assessment of its applicability using a diverse set of prompt examples drawn from sources on the web and from the DevGPT dataset~\cite{xiao2024devgpt}. Selected examples are presented in Table~\ref{tab:prompt-examples}. In addition, we analyzed a sample of 359 prompts extracted from DevGPT (\textasciitilde2\% of the prompts in DevGPT) and examined the extent to which prompt content could be mapped to the dimensions of the Prompt Triangle.

To obtain the sample, we randomly extracted initial prompts from conversations contained in \texttt{20230803\_095317\_commit\_sharings.json} and \texttt{20231012\_233628\_pr\_sharings.json}. We focused exclusively on initial prompts, as the objective at this stage was not to investigate prompt evolution during interactions, but rather to assess whether the Prompt Triangle dimensions are observable in standalone prompt formulations. Prompt content was categorized using Claude Sonnet 4.6 and subsequently reviewed manually to verify consistency. All prompts and analysis artifacts are available in the replication package~\cite{anonymous_2026_18713272}.

Our analysis shown in Table~\ref{tab:compact_analysis} demonstrates a strong tendency toward structured, multi-component prompts. Requirements (i.e., \emph{Functionality and Quality}) appeared in 80.8\% of prompts, establishing them as the foundational element of prompt construction. The predominant pattern ($n=247$, 68.8\%) incorporated all three components, indicating that users frequently provide comprehensive specifications of the solution. \emph{General Solutions} (79.7\%) are more commonly included than \emph{Specific Solutions} (69.4\%), suggesting that users prefer constraining the solution while preserving implementation flexibility. Co-occurrence analysis shows that requirements with at least one solution component accounted for all but two prompts. This asymmetry supports a way of thinking where specifying intent precedes the prescription of a solution. The secondary pattern of requirement-plus-general-solution prompts ($n=39$, 10.9\%) further supports a two-stage refinement process from ``what'' to ``how'' that may optionally extend to specific implementation details. Around 20\% of prompts were not applicable as they did not relate to coding (e.g., ``translate to Spanish'').

\begin{table}[tb]
\centering
\caption{Prompt Component Analysis (N=359)}
\label{tab:compact_analysis}
\begin{tabular}{lcccccc}
\toprule
& \multicolumn{3}{c}{\textbf{Component Present}} \\
\cmidrule(lr){2-4}
\textbf{Pattern} & \textbf{Req} & \textbf{Gen} & \textbf{Spec} & \textbf{n} & \textbf{\%} & \textbf{\% (excl.\ N/A)} \\
\midrule
All Three & \checkmark & \checkmark & \checkmark & 247 & 68.8 & 85.2 \\
Req + Gen & \checkmark & \checkmark &  & 39 & 10.9 & 13.4 \\
Req + Spec & \checkmark &  & \checkmark & 2 & 0.6 & 0.7 \\
Gen + Spec &  & \checkmark & \checkmark & 0 & 0 & 0 \\
Req Only & \checkmark &  &  & 2 & 0.6 & 0.7 \\
Gen Only &  & \checkmark &  & 0 & 0 & 0 \\
Spec Only &  &  & \checkmark & 0 & 0 & 0 \\
Not Applicable &  &  &  & 69 & 19.2 & --- \\
\midrule
\textbf{Total} & \textbf{290} & \textbf{286} & \textbf{249} & \textbf{359} & \textbf{100} & \textbf{100} \\
\textbf{(\%)} & \textbf{(80.8)} & \textbf{(79.7)} & \textbf{(69.4)} & & & \\
\bottomrule
\end{tabular}
\end{table}

\begin{table*}
\begin{threeparttable}
\centering
\caption{Example prompts with classification of the information contained in them to the components of the prompt triangle. More examples can be found in the supplementary material~\cite{anonymous_2026_18713272}.}
\label{tab:prompt-examples}
\footnotesize
\begin{tabular}{@{}p{5.2cm}P{3.2cm}P{2cm}P{3cm}@{}}
\toprule
\textbf{Prompt} & \textbf{Functionality and Quality} & \textbf{General Solution} & \textbf{Specific Solution} \\ 
\midrule Write a Python implementation of merge sort optimized for memory efficiency with time complexity analysis and error handling for edge cases including empty arrays.\tnote{1} & Sort input values; Provide time complexity analysis; Handle error/edge cases; Optimized for memory efficiency & Use merge sort algorithm; Implement in Python & \\
\midrule Create a TypeScript function that validates email addresses with the following requirements: Must be RFC 5322 compliant Rejects disposable email domains Returns detailed error messages. Example of expected function signature and usage: function validateEmail(email: string): \{ isValid: boolean; message: string \}\tnote{1} & Validate input; RFC 5322 compliance; Reject disposable email domains; Return detailed error messages & Use TypeScript & validateEmail(email: string) \\
\midrule I want to enable custom, unobtrusive update notifications and installs by customizing SPUUserDriver. Let's start by planning the custom UI we'll need. We'll ONLY work on UI. Create a plan for creating SwiftUI views that can show the various states that are required by SPUUserDriver. I think the best place for these to show up is in the macOS window titlebars on the top-right. Create a plan to put it there. Consult the oracle.\tnote{2} & Enable custom update notifications; Only UI work & & Put UI in macOS window titlebars; Use SwiftUI views \\
\midrule Generate a Python function using pandas to: Read a CSV file. Drop null values. Convert a date column to datetime format. Group data by category and calculate average values.\tnote{3} & Read CSV file; Drop null values; Convert date column; Group data by category; Calculate average values & Use pandas library; Implement in Python & Read with pandas.read\_csv \\ 
\midrule Write a FastAPI endpoint that: Accepts a POST request with JSON data (\{name: str, age: int\}). Validates the input. Returns a success message with the received data.\tnote{3} & Validate input; Return success message & Use FastAPI framework; Implement in Python & Define a POST endpoint; Accept JSON with \{name: str, age: int\} \\
\midrule Create a Python script using BeautifulSoup to: Fetch HTML from 'example.com/news'. Extract all headlines (h2 tags). Export results to a JSON file.\tnote{3} & Fetch HTML; Extract data from HTML; Export results to JSON & Use BeautifulSoup library; Implement in Python & Extract all <h2> tags \\
\bottomrule 
\end{tabular}
\begin{tablenotes}
    \item [1] \url{https://margabagus.com/prompt-engineering-code-generation-practices/}
    \item [2] \url{https://mitchellh.com/writing/non-trivial-vibing}
    \item [3] \url{https://zencoder.ai/blog/vibe-coding-prompts}
\end{tablenotes}
\end{threeparttable}
\end{table*}

\subsection{Prompting as a Tool for Requirements Validation and Verification} 
Requirements Engineering traditionally distinguishes between \emph{validation} (ensuring we are building the right thing) and \emph{verification} (ensuring we are building it right). In the context of AI-assisted chat-based development, structured prompts offer a novel mechanism to support both activities. By explicitly articulating \emph{Functionality and Quality}, \emph{General Solutions}, and \emph{Specific Solutions}, prompts act as evolving artifacts of stakeholder intent and technical interpretation. Requirement \emph{validation} occurs as the functionality component is refined in light of feedback and gradually reconciled with specific solutions, ultimately aligning the prompt with stakeholder needs. Meanwhile, requirement \emph{verification} is observable in the iterative interactions and adjustments between the three prompt components, which mirror the ongoing process of checking that the system is being built correctly. Thus, prompts not only guide the AI in code generation, they serve as dynamic roadmaps for both validating what should be built and verifying how it is implemented.

This bears similarities to classical requirement validation and verification techniques. Atoum et al.'s recent systematic review~\cite{9558838} categorizes these into six approaches. Our method resembles prototyping, inspection, and testing-oriented techniques. 

In terms of validation by \emph{inspection}, the \emph{Functionality and Quality} part of the Prompt Triangle serves a similar role to traditional requirements documents subjected to inspection. However, there are two distinctions: First of all, rather than discrete review meetings, the approach enables continuous, AI-mediated inspection through the iterative refinement between \emph{Functionality and Quality} and \emph{Specific Solutions}. The AI acts as an automated consistency analyzer that surfaces conflicts when it cannot reconcile high-level intent with implementation details, though the effectiveness of this detection depends on the AI's interpretive capabilities and how explicit the formulation in the prompt is. Second of all, we expect the \emph{Functionality and Quality} part of a prompt to be much more limited than a requirements document. An important part of inspections is the relationships between requirements~\cite{nuseibeh1994framework}. We do not expect such relationships to become visible in individual prompts, but potentially as part of longer conversations.

\emph{Prototyping} as a requirement validation technique was introduced by Boehm in 1984~\cite{boehm1984verifying}. In prototyping, working models of the software are created to validate that requirements meet stakeholder needs. One of the strengths of the Prompt Triangle is that it separates the solution components from the \emph{Functionality and Quality} parts, thus enabling rapid iteration between \emph{Functionality and Quality} and multiple solution formulations. Unlike traditional prototyping, which requires substantial development effort to explore alternative designs, the Prompt Triangle allows developers to experiment with different approaches in the \emph{General Solutions} and \emph{Specific Solutions} components while holding Functionality constant, generating executable prototypes within seconds rather than days~\cite{borg2025vibe}. When the AI fails to generate satisfactory code when a \emph{Specific Solution} is provided, this failure signals either requirement ambiguity or an incorrect solution approach. Notably, this mechanism also works in reverse: when the AI successfully generates code that nonetheless fails to meet stakeholder needs upon review, this reveals gaps in the \emph{Functionality and Quality} rather than implementation errors\,--\,a form of validation through executable prototyping.

In verification via \emph{testing-oriented approaches}, prompts inherently produce testable artifacts. By maintaining explicit traceability between \emph{Functionality and Quality} (what to test) and \emph{Specific Solutions} (how it is implemented) within the same artifact, the Prompt Triangle enables test generation that simultaneously validates stakeholder intent and verifies implementation correctness. However, the quality of generated tests depends critically on whether the AI grasps the essential test scenarios implied by \emph{Functionality and Quality} and the edge cases introduced by the \emph{Specific Solutions}. This dual focus is difficult to achieve when requirements and code exist as separate artifacts. In many cases, developers instruct AIs to generate test cases for each new functionality automatically. Research on automated test case generation with LLMs is just emerging (see, e.g., \cite{schafer2023empirical}), but combining requirement and solution approaches in a prompt can help guide the generation towards more targeted tests of the technical solution.

\subsection{Prompt Patterns and Prompt Iterations}
Recent studies analyze prompt patterns, but none treat prompts as blends of requirements and solutions.
DiCuffa et al.~\cite{dicuffa2025exploringpromptpatternsaiassisted} identify seven patterns, such as \emph{persona} (``you are'', ``pretend to be'') or \emph{recipe} (``step-by-step'', ``guide''), focusing on structural conventions rather than semantic content.
Siddiq et al.~\cite{siddiq2024fault} investigate quality issues with developer prompts, finding they often suffer from ambiguity and insufficient context for producing high-quality output. While these anti-patterns are complementary to our findings, we believe the Prompt Triangle can alleviate some of the issues Siddiq et al.\ identified by making the prompt components explicit and providing a way to reason about their balance.
Della Porta et al.~\cite{della2025prompt} analyze how prompt patterns affect code quality across three dimensions: meta-prompting (zero-shot and few-shot), chain-of-thought, and personas. They find no statistically significant differences between combinations, but like others, do not examine prompt content systematically.
Huang et al.~\cite{huang2025prompt} specifically address prompt engineering for requirements engineering, exploring techniques including multimodal prompting (e.g., UI mockups) and self-reflection prompts where LLMs provide confidence values. They propose using generated code to validate requirements, yet do not distinguish between requirement and solution elements within prompts themselves.
Beyond individual prompts, Robino~\cite{robino2025conversation} proposes ``conversation routines'' to structure extended human-LLM dialogues through system prompts, applicable to software development contexts.
Our prompt triangle provides the first systematic model decomposing prompts into requirement (Functionality and Quality) and solution (General, Specific) dimensions, enabling analysis of how these elements evolve through iterative refinement and therefore significantly goes beyond the state of the art in terms of analytical powers and the ability to craft high-quality prompts.

\section{Hypotheses}

The Prompt Triangle decomposes prompts into three complementary dimensions that characterize the information developers communicate when interacting with AI coding assistants: \emph{Functionality and Quality}, \emph{General Solutions}, and \emph{Specific Solutions}. While this decomposition helps explain \emph{what} information is present in prompts, understanding prompts as requirements artifacts also requires explaining \emph{how} these components are used throughout the development process.

As discussed in the previous sections, prompts increasingly resemble evolving software artifacts that support activities traditionally associated with Requirements Engineering (RE). Following established RE practice~\cite{nuseibeh2000requirements}, we distinguish between three fundamental activities: \emph{validation}, concerned with ensuring that the right system is being built; \emph{verification}, concerned with ensuring that the system is built correctly; and \emph{solution generation}, concerned with exploring and realizing implementation alternatives. While the Prompt Triangle captures the semantic structure of prompt content, these RE activities characterize the role such content plays during development interactions.

Reflecting on the relationships between Prompt Triangle components and RE activities raises several questions concerning how prompts evolve, how developers compose them, and how prompt composition may influence resulting software quality. Based on the conceptual arguments developed throughout this paper and prior observations of iterative AI-assisted development~\cite{barke2023grounded,shalini2026exploring}, we formulate the following hypotheses:

\begin{description}

\item[H1 (Prompt Evolution)]
During iterative AI-assisted development, prompt content gradually shifts from emphasizing \textbf{Functionality and Quality} toward increasingly detailed \textbf{General Solutions} and \textbf{Specific Solutions} information.

\item[H2 (Developer-specific Prompt Composition)]
The composition of Prompt Triangle components differs according to developer characteristics, such as programming experience and domain familiarity. In particular, developers with greater experience are expected to provide proportionally more \textbf{Specific Solutions} information.

\item[H3 (Prompt Components and RE Activities)]
Prompt Triangle components are associated with different RE activities. Specifically, \textbf{Functionality and Quality} information is expected to occur more frequently during validation-oriented interactions, whereas \textbf{Specific Solutions} information is expected to occur more frequently during verification-oriented interactions.

\item[H4 (Prompt Alignment and Development Outcomes)]
\sloppy AI-assisted development interactions exhibiting stronger alignment between Prompt Triangle components and their corresponding RE activities are expected to produce higher software quality outcomes.

\end{description}

Together, these hypotheses define an empirical research agenda for studying prompts as evolving requirements artifacts. H1 and H2 focus on understanding how prompt composition emerges and changes across interactions and users, whereas H3 and H4 investigate whether these structures influence development processes and software outcomes.

\section{Discussion}
The power of chat-based interfaces is most obvious in vibe coding.
Meske et al.~\cite{meske2025vibecodingreconfigurationintent} define vibe coding as the culmination of a historical process of intent mediation: ``the level of abstraction required to implement solutions has evolved from hardware manipulation through conceptual modeling to prompting, shifting the mediation of developer intent from deterministic instruction to probabilistic inference.'' To guide this inference, ``goal-oriented intent expression'' via prompts is necessary. Crucially, Meske et al.~consider actual implementation details irrelevant. However, in the chat-based coding prompts we have studied, solution details as captured in the Prompt Triangle are almost always present.

This discrepancy suggests either that current practice remains anchored to solution-level control due to developer trust issues or LLM limitations, or that Meske et al.'s vision represents an idealized future state not yet achievable. Our Prompt Triangle embraces this reality by explicitly accommodating solution guidance alongside functional intent, rather than treating solution details as impurities to eliminate. As LLMs mature, we hypothesize the relative weight of Specific Solutions may decrease, but the tripartite structure will remain relevant as developers will always need some architectural control. Our work on H1 and H2 will identify how developers evolve prompts during their work, while H3 and H4 will reveal how requirements and solutions co-evolve and whether this strategy benefits code quality.

Taking this further, OpenAI's Sean Grove proposes that prompts become executable specifications, arguing future programmers will write elaborate specifications that generative AIs turn into complex programs\footnote{\url{https://www.youtube.com/watch?v=8rABwKRsec4}}. Counter-arguments exist: executable specifications have long existed~\cite{hayes1989specifications,fuchs1992specifications}, and making specifications fully executable constitutes programming itself. Moreover, unlike formal programming languages, natural language is ambiguous.
Jackson considers this ambiguity a strength rather than weakness for natural language reasoning in AI~\cite{jackson2020understanding}. LLMs arguably realize his vision by connecting natural language to vector space concepts. However, Jackson envisioned reasoning AIs, not code generators: the ability to ``understand the world'' via natural language does not directly translate to reproducibly creating programs fulfilling user needs. Much requirements engineering work addresses disambiguating unclear requirements~\cite{khalil2025detecting}.

Our work addresses this conflict directly: we study how requirements are captured in prompts with all their ambiguity and how engineers constrain generation by providing solution approaches. We believe prompt requirements must evolve through disambiguation and scaffolding. H1 and H3 will illuminate this process, while H4 will reveal the impact on code quality when validation and disambiguation happen early.

The tool landscape validates our premise. Kiro\footnote{\url{https://kiro.dev/}} promises to turn ``your prompt into clear requirements, system design, and discrete tasks.'' Here, prompts come first and requirements after\,--\,validating that requirements are inherent in prompts alongside solution aspects. Kiro's value proposition is extracting and disambiguating these elements.

\section{Plans for Empirical Validation}
We intend to employ a \emph{multi-method data collection strategy} to construct a comprehensive corpus of developer–AI interactions. Our approach combines three complementary sources of data: (i) controlled experiments, (ii) community-sourced prompt exports, and (iii) mining of publicly available datasets. Figure~\ref{fig:methodology} shows the complete vision for this study. It is described in more detail in a registered report~\cite{chakraborty_2026_21351049}.

Our triangulated design balances internal and ecological validity through three complementary data sources. Following Colnet et al.~\cite{colnet2024causal}, who review methods for combining randomized trials and observational studies, our controlled experiment ($n=30$) provides unconfounded causal estimates with full quality measures, while community uploads (target $n \geq 150$) and mined public datasets assess generalizability through observational data. Confirmatory hypothesis testing (H1--H4) is based on controlled experiment data; community uploads serve as co-primary replication data for H2 and supplementary validation for H1 and H3; mined datasets serve an explicitly exploratory role for pattern characterization only.
Table~\ref{tab:collection-methods} summarizes the collection approaches and their contributions to the corpus.

\begin{table*}[tb]
\centering
\caption{Comparison of data collection methods}
\label{tab:collection-methods}
\begin{tabular}{@{}P{2.1cm}P{2cm}P{5.2cm}P{3.4cm}@{}}
\toprule
\textbf{Method} & \textbf{Data Type} & \textbf{Strengths} & \textbf{Limitations} \\
\midrule
Controlled Experiments & Structured, curated & Complete metadata, ground truth requirements, comparable tasks & Limited scale, potential observer effects \\
Community Uploads & Real-world, diverse & Authentic usage, varied contexts, scalable & Incomplete metadata, self-selection bias \\
Public Mining & Opportunistic & No recruitment overhead, naturally shared & Sparse, variable quality \\
\bottomrule
\end{tabular}
\end{table*}

\begin{figure*}[t]
\centering
\resizebox{\textwidth}{!}{
\begin{tikzpicture}[
node distance=1.4cm and 2.2cm,
box/.style={
draw,
rounded corners,
align=center,
minimum width=3.5cm,
minimum height=1cm,
font=\small
},
arrow/.style={->, thick}
]

\node[box] (exp) {
\textbf{Controlled Experiments}\\
$N=30$ developers\\
4 coding tasks\\
Chat logs, code, recordings
};

\node[box,right=of exp] (community) {
\textbf{Community Uploads}\\
$n\geq150$ conversations\\
Voluntary IDE exports\\
Demographics + context
};

\node[box,right=of community] (mining) {
\textbf{Public Mining}\\
$\geq200$ conversations\\
GitHub, Reddit, Blogs\\
DevGPT and social platforms
};

\node[box,below=.6cm of community] (processing) {
\textbf{Unified Processing Pipeline}\\
Format normalization $\rightarrow$
Quality checks $\rightarrow$
PII removal $\rightarrow$
Annotation
};

\node[box,below=.4cm of processing] (annotation) {
\textbf{Prompt Analysis}\\
Prompt Triangle coding\\
(F\&Q, General Solutions, Specific Solutions)\\
RE activity coding + metadata
};

\node[box,below=.4cm of annotation] (analysis) {
\textbf{Hypothesis Testing}\\
H1: Prompt evolution\\
H2: User characteristics\\
H3: RE activity alignment\\
H4: Temporal alignment and quality
};

\draw[arrow] (exp) -- (processing);
\draw[arrow] (community) -- (processing);
\draw[arrow] (mining) -- (processing);

\draw[arrow] (processing) -- (annotation);
\draw[arrow] (annotation) -- (analysis);

\end{tikzpicture}
}
\caption{Overview of the mixed-method study design. Three complementary data sources (controlled experiments, community-sourced uploads, and publicly mined conversations) are integrated through a unified processing and annotation pipeline to investigate the proposed hypotheses.}
\label{fig:methodology}
\end{figure*}

\section{Conclusion}
Chat-based development shifts the developer's role from writing code toward crafting prompts, positioning prompts as artifacts that increasingly blend requirements with solution guidance. The Prompt Triangle provides a structured lens to characterize prompt content, enabling the analysis of prompt composition, evolution, and its relationship to requirements engineering activities and software outcomes. Based on this perspective, we formulate hypotheses on prompt evolution, developer strategies, and software quality. Future work will investigate these hypotheses to derive evidence-based practices and tool support for requirements-aware prompt engineering. We encourage the RE community to view prompting as a central concern for the future of software engineering.

\section*{Acknowledgments}
Perplexity.ai and Claude Sonnet 4.5 and 4.6 were used to critique drafts of the manuscript, refine hypotheses, and elicit feedback on different aspects of the conceptual model. Claude Sonnet 4.6 was used to generate the scripts we used to prepare the dataset, to classify information in prompts into the three areas of the triangle, and to generate the corresponding tables. In addition, it created some of the text passages in this work.

\section*{Data Availability Statement}
The following materials are available in the supplementary material~\cite{anonymous_2026_18713272}: (1) a small set of prompts from sources on the web that were manually classified; (2) a set of 359 prompts randomly extracted from the DevGPT dataset and classified by Claude Sonnet 4.6; (3) a set of scripts to extract, classify, and visualize prompts from DevGPT.

\bibliographystyle{plainurl}
\bibliography{ref}

@article{ray2025review,
  title={A Review on Vibe Coding: Fundamentals, State-of-the-art, Challenges and Future Directions},
  author={Ray, Partha Pratim},
  journal={Authorea Preprints},
  year={2025},
  publisher={Authorea}
}

@article{ronanki2025prompt,
  title={Prompt Engineering Guidelines for Using Large Language Models in Requirements Engineering},
  author={Ronanki, Krishna and Arvidsson, Simon and Axell, Johan},
  journal={arXiv preprint arXiv:2507.03405},
  year={2025}
}

@inproceedings{huang2025prompt,
  title={Prompt engineering for requirements engineering: A literature review and roadmap},
  author={Huang, Kaicheng and Wang, Fanyu and Huang, Yutan and Arora, Chetan},
  booktitle={RE Workshops},
  pages={548--557},
  year={2025},
  organization={IEEE}
}

@inproceedings{reynolds2021prompt,
  title={Prompt programming for large language models: Beyond the few-shot paradigm},
  author={Reynolds, Laria and McDonell, Kyle},
  booktitle={CHI},
  pages={1--7},
  year={2021}
}

@article{chen2023unleashing,
  title={Unleashing the potential of prompt engineering in large language models: a comprehensive review},
  author={Chen, Banghao and Zhang, Zhaofeng and Langren{\'e}, Nicolas and Zhu, Shengxin},
  journal={arXiv preprint arXiv:2310.14735},
  year={2023}
}

@inproceedings{vogelsang2024prompting,
  title={Prompting the Future: Integrating Generative LLMs and Requirements Engineering.},
  author={Vogelsang, Andreas},
  booktitle={REFSQ Workshops},
  year={2024}
}

@inproceedings{kruger2023intentions,
author = {Kr\"{u}ger, Jacob and Li, Yi and Zhu, Chenguang and Chechik, Marsha and Berger, Thorsten and Rubin, Julia},
title = {A Vision on Intentions in Software Engineering},
year = {2023},
isbn = {9798400703270},
publisher = {Association for Computing Machinery},
address = {New York, NY, USA},
url = {https://doi.org/10.1145/3611643.3613087},
doi = {10.1145/3611643.3613087},
booktitle = {FSE/ESEM},
pages = {2117–2121},
numpages = {5},
location = {San Francisco, CA, USA},
series = {ESEC/FSE 2023}
}

@article{meske2025vibecodingreconfigurationintent,
  title={Vibe coding as a reconfiguration of intent mediation in software development: Definition, implications, and research agenda},
  author={Meske, Christian and Hermanns, Tobias and Von der Weiden, Esther and Loser, Kai-Uwe and Berger, Thorsten},
  journal={IEEE Access},
  volume={13},
  pages={213242--213259},
  year={2025},
  publisher={IEEE}
}

@article{sergeyuk2025using,
  title={Using AI-based coding assistants in practice: State of affairs, perceptions, and ways forward},
  author={Sergeyuk, Agnia and Golubev, Yaroslav and Bryksin, Timofey and Ahmed, Iftekhar},
  journal={IST},
  volume={178},
  pages={107610},
  year={2025},
  publisher={Elsevier}
}

@article{hayes1989specifications,
  title={Specifications are not (necessarily) executable},
  author={Hayes, Ian James and Jones, Cliff B},
  journal={Software Engineering Journal},
  volume={4},
  number={6},
  pages={330--339},
  year={1989},
  publisher={IET}
}

@article{fuchs1992specifications,
  title={Specifications are (preferably) executable},
  author={Fuchs, Norbert E},
  journal={Software Engineering Journal},
  volume={7},
  number={5},
  pages={323--334},
  year={1992},
  publisher={IET}
}

@article{robeyns2025self,
  title={A self-improving coding agent},
  author={Robeyns, Maxime and Szummer, Martin and Aitchison, Laurence},
  journal={arXiv preprint arXiv:2504.15228},
  year={2025}
}

@article{wang2025ai,
  title={AI Agentic Programming: A Survey of Techniques, Challenges, and Opportunities},
  author={Wang, Huanting and Gong, Jingzhi and Zhang, Huawei and Wang, Zheng},
  journal={arXiv preprint arXiv:2508.11126},
  year={2025}
}

@article{jackson2020understanding,
  title={Understanding understanding and ambiguity in natural language},
  author={Jackson, Philip},
  journal={Procedia Computer Science},
  volume={169},
  pages={209--225},
  year={2020},
  publisher={Elsevier}
}

@inproceedings{khalil2025detecting,
  title={Detecting Cross-Domain Ambiguity In Requirements Through Natural Language Processing, A Systematic Literature Review},
  author={Khalil, Ibrahim and Ahmad, Israr and Rasheed, Uzair and Butt, Wasi Haider and Anwaar, Zaeem},
  booktitle={ICACS},
  year={2025},
  organization={IEEE}
}

@article{wei2022chain,
  title={Chain-of-thought prompting elicits reasoning in large language models},
  author={Wei, Jason and Wang, Xuezhi and Schuurmans, Dale and Bosma, Maarten and Xia, Fei and Chi, Ed and Le, Quoc V and Zhou, Denny and others},
  journal={Advances in neural information processing systems},
  volume={35},
  pages={24824--24837},
  year={2022}
}

@article{barke2023grounded,
  title={Grounded copilot: How programmers interact with code-generating models},
  author={Barke, Shraddha and James, Michael B and Polikarpova, Nadia},
  journal={OOPSLA},
  volume={7},
  pages={85--111},
  year={2023},
  publisher={ACM New York, NY, USA}
}

@inproceedings{vaithilingam2022expectation,
  title={Expectation vs. experience: Evaluating the usability of code generation tools powered by large language models},
  author={Vaithilingam, Priyan and Zhang, Tianyi and Glassman, Elena L},
  booktitle={CHI},
  pages={1--7},
  year={2022}
}

@inproceedings{dicuffa2025exploringpromptpatternsaiassisted,
  title={Exploring Prompt Patterns in AI-Assisted Code Generation: Towards Faster and More Effective Developer-AI Collaboration},
  author={DiCuffa, Sophia and Zambrana, Amanda and Yadav, Priyanshi and Madiraju, Sashidhar and Suman, Khushi and AlOmar, Eman Abdullah},
  booktitle={ICMI},
  pages={1--7},
  year={2025},
  organization={IEEE}
}

@inproceedings{siddiq2024fault,
  title={The fault in our stars: Quality assessment of code generation benchmarks},
  author={Siddiq, Mohammed Latif and Dristi, Simantika and Saha, Joy and Santos, Joanna CS},
  booktitle={SCAM},
  pages={201--212},
  year={2024},
  organization={IEEE}
}

@inproceedings{della2025prompt,
  title={Do Prompt Patterns Affect Code Quality? A First Empirical Assessment of ChatGPT-Generated Code},
  author={Della Porta, Antonio and Lambiase, Stefano and Palomba, Fabio},
  booktitle={EASE},
  pages={181--192},
  year={2025}
}

@inproceedings{xiao2024devgpt,
  title={Devgpt: Studying developer-chatgpt conversations},
  author={Xiao, Tao and Treude, Christoph and Hata, Hideaki and Matsumoto, Kenichi},
  booktitle={MSR},
  pages={227--230},
  year={2024}
}

@article{robino2025conversation,
  title={Conversation routines: A prompt engineering framework for task-oriented dialog systems},
  author={Robino, Giorgio},
  journal={arXiv preprint arXiv:2501.11613},
  year={2025}
}

@inproceedings{VillamizarFKVM25,
  author       = {Hugo Villamizar and
                  Jannik Fischbach and
                  Alexander Korn and
                  Andreas Vogelsang and
                  Daniel M{\'{e}}ndez},
  title        = {Prompts as Software Engineering Artifacts: {A} Research Agenda and
                  Preliminary Findings},
  booktitle    = {{PROFES}},
  series       = {LNCS},
  volume       = {16361},
  pages        = {470--478},
  publisher    = {Springer},
  year         = {2025},
  url          = {https://doi.org/10.1007/978-3-032-12089-2\_32},
  doi          = {10.1007/978-3-032-12089-2\_32},
  bibsource    = {dblp computer science bibliography, https://dblp.org}
}

@ARTICLE{9558838,
  author={Atoum, Issa and Baklizi, Mahmoud Khalid and Alsmadi, Izzat and Otoom, Ahmed Ali and Alhersh, Taha and Ababneh, Jafar and Almalki, Jameel and Alshahrani, Saeed Masoud},
  journal={IEEE Access}, 
  title={Challenges of Software Requirements Quality Assurance and Validation: A Systematic Literature Review}, 
  year={2021},
  volume={9},
  number={},
  pages={137613-137634},
  doi={10.1109/ACCESS.2021.3117989}}

@ARTICLE{nuseibeh1994framework,
  author={Nuseibeh, B. and Kramer, J. and Finkelstein, A.},
  journal={TSE}, 
  title={A framework for expressing the relationships between multiple views in requirements specification}, 
  year={1994},
  volume={20},
  number={10},
  pages={760-773},
  doi={10.1109/32.328995}}

@ARTICLE{boehm1984verifying,
  author={Boehm, B.W.},
  journal={IEEE Software}, 
  title={Verifying and Validating Software Requirements and Design Specifications}, 
  year={1984},
  volume={1},
  number={1},
  pages={75-88},
  doi={10.1109/MS.1984.233702}}

@article{schafer2023empirical,
  title={An empirical evaluation of using large language models for automated unit test generation},
  author={Sch{\"a}fer, Max and Nadi, Sarah and Eghbali, Aryaz and Tip, Frank},
  journal={TSE},
  volume={50},
  number={1},
  pages={85--105},
  year={2023},
  publisher={IEEE}
}

@article{borg2025vibe,
  title={Vibe coding and the new prototyping playbook},
  author={Borg, Markus and Bjarnason, Elizabeth and Hedin, Fabian},
  journal={IEEE Software},
  volume={42},
  number={6},
  pages={12--16},
  year={2025},
  publisher={IEEE}
}

@inproceedings{shalini2026exploring,
    author = {Chakraborty, Shalini and Steghöfer, Jan-Philipp},
    title = {Exploring Prompts as Mixed Requirements and Solutions Artifacts},
    booktitle = {{ICSE} Companion},
    year = 2026
}

@misc{anonymous_2026_18713272,
  author       = {Chakraborty, Shalini and Steghöfer, Jan-Philipp},
  title        = {{Dataset for ``Prompts Blend Requirements and
                   Solutions: From Intent to Implementation''}},
  month        = may,
  year         = 2026,
  publisher    = {Zenodo},
  doi          = {10.5281/zenodo.20329289},
  url          = {https://doi.org/10.5281/zenodo.20329289},
}

@inproceedings{nuseibeh2000requirements,
  title={Requirements engineering: a roadmap},
  author={Nuseibeh, Bashar and Easterbrook, Steve},
  booktitle={Proceedings of the Conference on the Future of Software Engineering},
  pages={35--46},
  year={2000}
}

@article{colnet2024causal,
  title={Causal inference methods for combining randomized trials and observational studies: a review},
  author={Colnet, B{\'e}n{\'e}dicte and Mayer, Imke and Chen, Guanhua and Dieng, Awa and Li, Ruohong and Varoquaux, Ga{\"e}l and Vert, Jean-Philippe and Josse, Julie and Yang, Shu},
  journal={Statistical Science},
  volume={39},
  number={1},
  pages={165--191},
  year={2024},
  publisher={Institute of Mathematical Statistics}
}

@misc{chakraborty_2026_21351049,
  author       = {Chakraborty, Shalini and
                  Steghöfer, Jan-Philipp},
  title        = {The Prompt Triangle: A Registered Report on
                   Prompts as Hybrid Artifacts},
  booktitle    = {20th Int. Symposium on Empirical Software Engineering and Measurement (ESEM 2026)},
  month        = jul,
  year         = 2026,
  publisher    = {Zenodo},
  doi          = {10.5281/zenodo.21351049},
  url          = {https://doi.org/10.5281/zenodo.21351049},
}

\end{document}